\documentclass[journal]{IEEEtran}
\usepackage{graphicx,cite,epsfig,amssymb,amsmath}
\usepackage{pifont}
\usepackage{stmaryrd}
\usepackage{bbding}
\usepackage{algorithm}
\usepackage{algorithmic}
\usepackage{mathrsfs}
\usepackage{latexsym}
\usepackage{indentfirst}
\usepackage{fmtcount}
\usepackage{cite}
\usepackage{float}
\usepackage{morefloats}

\usepackage{caption}
\usepackage{subcaption}

\usepackage{color}
\usepackage{mathtools}
\usepackage{xfrac}

\usepackage{array,colortbl,multirow,multicol,booktabs,ctable,bigstrut}

\usepackage[
top    = 0.75in,
bottom = 0.75in,
left   = 0.75in,
right  = 0.75in]{geometry}

\begin{document}
%
\title{Reed-Muller Sequences for 5G Grant-free Massive Access}
%
%
%

\author{
    \IEEEauthorblockN{Huazi~Zhang\IEEEauthorrefmark{1}, Rong~Li\IEEEauthorrefmark{1}, Jun~Wang\IEEEauthorrefmark{1}, Yan~Chen\IEEEauthorrefmark{1} and~Zhaoyang~Zhang\IEEEauthorrefmark{2}}\\
    \IEEEauthorblockA{\IEEEauthorrefmark{1}Huawei Technologies Co. Ltd.}\\
    \IEEEauthorblockA{\IEEEauthorrefmark{2}College of Information Science \& Electronic Engineering, Zhejiang University, China}\\
    Email: zhanghuazi@huawei.com, ning\_ming@zju.edu.cn
}
\maketitle

\begin{abstract}
We propose to use second order Reed-Muller (RM) sequence for user identification in 5G grant-free access. The benefits of RM sequences mainly lie in two folds, (i) support of much larger user space, hence lower collision probability and (ii) lower detection complexity. These two features are essential to meet the massive connectivity ($10^7$ links/km$^2$), ultra-reliable and low-latency requirements in 5G, e.g., one-shot transmission ($\leq 1$ms) with $\leq 10^{-4}$ packet error rate. However, the non-orthogonality introduced during sequence space expansion leads to worse detection performance. In this paper, we propose a noise-resilient detection algorithm along with a layered sequence construction to meet the harsh requirements. Link-level simulations in both narrow-band and OFDM-based scenarios show that RM sequences are suitable for 5G.
\end{abstract}

\begin{IEEEkeywords}
5G, Grant-free access, Reed-Muller Sequences, Internet of Things (IoT).
\end{IEEEkeywords}

\IEEEpeerreviewmaketitle

\section{Introduction}
\subsection{Motivation}
We are ushering in the fifth-generation (5G) wireless communications \cite{5G:concept}. The rich diversity of applications are driving technologies towards not only higher bandwidth and throughput, but a variety of metrics. This application-driven network (ADN) vision will potentially revolutionize wireless networking from all aspects, including the physical layer. The use cases will be very different from the incumbent long term evolution (LTE). First, 5G should support \textit{massive connectivity} with a much larger number of devices, e.g., $10^7$ links/km$^2$. Second, in mission-critical scenarios, such as vehicular-to-vehicular networks, \textit{ultra-high reliability} and \textit{low latency}, e.g., $\leq 10^{-4}$ packet loss rate within $\leq 1$ms response time, should be supported \cite{3GPP:critical_communications} \cite{5G:features}.

Current wireless systems, such as 4G LTE and WiFi, are not designed to support the above-mentioned features. In LTE, scheduling is required to establish a connection between a user equipment (UE) and a base station (BS). In the context of massive connectivity and ultra-low-latency communications, scheduling has two weaknesses. First, short packet transmissions will be the dominant traffic pattern. In these cases, scheduling will lead to \textit{high signaling-to-data ratio} and low spectrum efficiency. Second, the extra round-trip delay time consumed by scheduling incurs \textit{unacceptable latency}. In the IEEE 802.11 standards, scheduling-free transmissions are allowed with the help of carrier sense multiple access with collision avoidance (CSMA/CA). However, it only support local area networking with a small number of users.

\subsection{Grant-free multiple access}
In order to fulfill the massive connectivity ($10^7$ links/km$^2$), ultra-high reliability ($\leq 10^{-4}$ packet loss rate) and ultra-low latency ($\leq1$ms response) promises in 5G IoT, a sparse code multiple access (SCMA) based uplink grant-free design \cite{SCMA:grant-free} is proposed to eliminate the scheduling procedure. In order to support low latency, the basic radio resource for grant-free transmission is a contention transmission unit (CTU), defined as a combination of time, frequency and pilot sequence. A user is allowed to transmit data on the CTUs immediately after packet arrival without waiting for a grant. During an uplink transmission, a UE transmits both a pilot sequence (or preamble) and data \textit{in one shot}, and the BS jointly decodes the data of all users from the superimposed signal. Since the length of the entire packet is usually very short, ultra-low latency can be achieved. The benefits of grant-free access are two-fold, much shorter access delay and lower overhead ratio.
\begin{figure}
\centering
    \includegraphics[width= 0.47\textwidth]{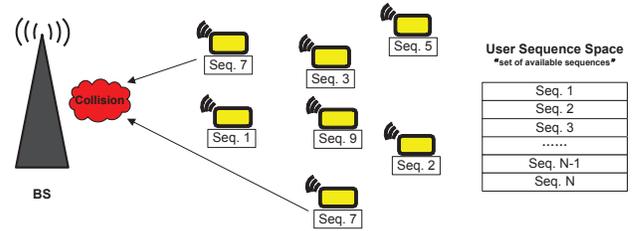}
    \caption{Collision in grant-free massive access.}
    \label{fig:grant_free}
\end{figure}

\begin{table}
  \centering
  \caption{Collision rate in contention-based access}
    \begin{tabular}{|c|c|c|c|}
    \hline
    \multirow{2}[4]{*}{Sequence Space } & \multicolumn{3}{c|}{Number of Active Users} \bigstrut\\
\cline{2-4}          & 2     & 4     & 6 \bigstrut\\
    \hline
    52 (LTE PRACH \cite{3GPP:PRACH}) & 0.0192 & 0.0566 & 0.0925 \bigstrut\\
    \hline
    16000 (5G, proposed) & $6 \times 10^{-5}$ & $1.8 \times 10^{-4}$ & $3 \times 10^{-4}$ \bigstrut\\
    \hline
    \end{tabular}%
  \label{tab:collisionRate}%
\end{table}%

However, the performance bottleneck of grant-free access turns out to be the ``collisions'' among users. As shown in Fig.~\ref{fig:grant_free}, each active user randomly chooses a sequence from a ``sequence space'', and all users simultaneously transmit on the same time-frequency resource block. There is a chance that two users choose the same sequence to access, resulting in a collision-incurred packet loss. As shown in Table~\ref{tab:collisionRate}, the collision rate is determined only by the size of sequence space and the number of simultaneous accessing users. As seen, a large sequence space is necessary for grant-free access.

The pilot sequence is of paramount importance here because it not only identifies a user but also enables channel estimation and indicates the SCMA codebook being used, all of which are necessary for a successful transmission. The current contention-based access scheme in LTE, such as the Physical Random Access Channel (PRACH), may not support massive connectivity mainly due to the fact that the user sequence being used (i.e. Zadoff-Chu (ZC) sequence). When the number of active users grows, picking from a small and fixed-sized sequence pool will inevitably incur high collision rate between users. According to LTE \cite{3GPP:PRACH}, the number of ZC sequences for contention-based random access is 52. As shown in Table~\ref{tab:collisionRate}, the collision rate is close to 0.1 when six users simultaneously access. One straightforward way to alleviate this issue is expanding the sequence space with longer ZC sequences. In practice, this may not be a wise option for grant-free access due to (i) high overhead ratio and (ii) high detection complexity.

We propose to use second-order Reed-Muller (RM) codes \cite{RM:a-fast} for grant-free massive access in 5G, both as sequences for user detection and as demodulation reference signal for channel estimation. RM sequences and its detection process has the following attractive features: (i) it can create a sequence space of orders of magnitudes larger than ZC with same-length sequences, (ii) in both small- and large-sized sequence spaces, the detection algorithm can be much faster than that of ZC sequences. To our best knowledge, although RM sequences have wide applications in image processing, its potential in wireless communications is yet to be exploited. In \cite{RM:neighbor}, full-duplex neighbor discovery is proposed in a fully distributed network, based on on-off RM sequences with erasures. In \cite{RM:CDMA}, RM sequences are used for downlink modulation to achieve a higher sum rate. In this paper, we focus on its application in 5G massive connectivity and ultra-low latency communications. Our contributions are summarized as follows:

\begin{enumerate}
\item It is the first to use Reed-Muller sequences for grant-free massive access. The huge RM sequence space can greatly reduce collisions between users during contention-based access. We illustrate a collision-detection tradeoff due to the non-orthogonality incurred during sequence space expansion, and propose a layered RM construction to reduce multi-user interference.
\item To cope with the inherent noisy nature of wireless channels, we made a variety of improvements in the detection algorithm. In particular, we shuffle over multiple orders to recover the columns of $P$ matrix which corresponds to the detected user, and propose two decision metrics to pick the most reliable one from a set of candidates. The proposed algorithm is shown to have significant performance gain.
\item We implemented Reed-Muller sequences in both narrow-band IoT and wideband OFDM-SCMA under realistic parameter setting. Both the dramatically increased sequence space (20x to 50x) and our noise-resilient detection algorithm have contributed to the significant performance gain in terms of collision rate, detection rate and block error rate (BLER).
\end{enumerate}

\section{Reed-Muller sequence for user identification}\label{section:RM}
According to the contention-based grant-free access, a number of users simultaneously transmit on a particular contention region. When using RM sequences $\phi_{P_l,b_l}$, the received aliased signal is
\begin{equation}\label{y}
y(t)=\sum_{l=1}^k h_l \phi_{P_l,b_l}(t) + n(t),
\end{equation}
where $k$ is the total number of active users, $\phi_{P_l,b_l}$ is the RM sequence of the $l$-th active user, $h_l$ is the channel between the $l$-th user and the BS, and $n(t)$ is white Gaussian noise. Our goal is to detect from the aliased signal all $k$ active users, recover the transmitted signal $\hat \phi$ and estimate the corresponding channel $\hat h$.

Reed-Muller (RM) sequences of length $2^m$ can create up to a $2^{m(r+2)}$-sized sequence space, and is parameterized by $(m,r)$. Given a user ID in a $C$-sized user space, we propose to construct RM sequences through the following steps:
\begin{enumerate}
\item Choose a user space of size $C \leq 2^{m(r+2)}$.
\item Convert a user ID $\in \{0,\cdots ,C-1\}$ to $m(r+2)$ bits in the binary form.
\item Map the $m(r+2)$-bit user ID to an $m \times m$-sized $P$ matrix and an $m$-length $b$ vector as follows:
\begin{itemize}
\item Take the least significant $m$ bits as the $b$ vector.
\item Take the rest $m(r+1)$ bits and evenly partition them into $r+1$ groups. The group containing the least significant $m$ bits are mapped to a $P_0$ matrix in the Kerdock set \cite{RM:construction}; the groups containing the higher bits are mapped to matrices $P_t, t \in \{1, \cdots, r\}$, each corresponding to a matrix in the Delsarte-Goethals $(m,t)$ set \cite{RM:construction}. Now we have $m+1$ matrices in total and sum them up in $GF(2)$, and obtained the $P$ matrix.
\end{itemize}
\item Construct the $2^m$-bit Reed-Muller sequence corresponding to the user ID as follows:
\begin{equation}\label{reed_function}
\phi_{P,b}(x)={\frac {(-1)^{weight(b)}} {\sqrt{2^m}}} i^{(2b+Px)^{T}x},
\end{equation}
where $x$ is an $m$-length binary vector, which indicates the index of the RM sequence ranging from 1 to $2^m$. The above function is called the second-order Reed-Muller function (see \cite{RM:construction} and the references therein).
\end{enumerate}

Reed-Muller sequences have the following properties:
\begin{enumerate}
\item RM sequence has length $2^m$ and all values are taken from $\{1,i,-1,-i\}$. The maximal value of $r$ is $\lfloor \frac{m-1}{2} \rfloor$, thus the $2^m$-length sequence can create up to a $2^{m(r+2)} = 2^{\frac{m(m+3)}{2}}$-sized sequence space to support a same-sized user space.
\item RM sequences are well structured. If we pointwise-multiply any $\phi_{P,b}$ with the conjugate of $\phi_{P,0}$, the result is a Walsh function determined by $b$. All $2^m$ Walsh functions form the rows of Hadamard matrix $H_m$, which is constructed in the following recursive fashion:
\begin{equation}
H_m = \begin{bmatrix}H_{m-1} & H_{m-1} \\ H_{m-1} & -H_{m-1} \end{bmatrix}.
\end{equation}
\item The $2^m$ RM sequences generated from the same $P$ are orthogonal. The two RM sequences generated from two distinct matrices $P$ and $Q$, $P,Q \in DG(m,r)$, have coherence \cite{RM:construction}
\begin{equation}
\mu^{P,Q} = \left\{\begin{matrix}
{\frac 1 {\sqrt {2^{m-2r}}}}, & 2^{m-2r} \quad times,\\
0, & 2^m-2^{m-2r} \quad times.
\end{matrix}\right.
\end{equation}
\end{enumerate}

In the context of grant-free massive access \cite{SCMA:grant-free}, we have both good news and bad news. The first good news, thanks to the first property, is that we can create a user space of much larger and flexible size. This feature may solve the user sequence resource scarcity problem under massive connectivity. At least, it offers huge flexibility to expand the user space when needed.

Another good news is a fast detection algorithm brought by the second property. Owing to the recursive structure of Walsh functions, we can determine an unknown $b$ by performing a fast Walsh-Hadamard transform, which only takes $O(n \log n)$ multiplications as compared with $O(n^2)$ for the correlation method. Moreover, since RM sequences only take values from $\{+1,-1,+i,-i\}$, the multiplications only requires flip of signs which is extremely simple. This feature alone provides tremendous complexity reduction than other sequences such as Zadoff–Chu (ZC) sequence in LTE. For example, for a 64-length sequence, RM requires only $\frac{1}{10}$ multiplications of ZC, and the complexity of each multiplication is \textit{negligible} compared with ZC.

The bad news is the non-orthogonality introduced during space expansion. The last property reveals to us the fundamental tradeoff between user space size (collision probability) and inter-user interference (detection performance). As we expand the user space, we have to compromise on the orthogonality and thus inevitably loose some detection performance.

\section{Collision-detection tradeoff and layered construction of RM sequences}
As aforementioned, there exists a fundamental tradeoff between better orthogonality and larger sequence space. In grant-free massive access, the former implies lower multi-user interference and thus detection performance; the latter is associated to collision rate.

Given the sequence size $C$ and active user number $k$, the collision rate drops as the sequence space expands
\begin{equation}
p^{col}(k,C)={\frac C k}\sum_{i=2}^k \left(iC_k^i\left(\frac 1 C\right)^i\left(\frac {C-1} C\right)^{k-i}\right)
\end{equation}

The detection performance depends on the specific sequences we use and the detection algorithm. The rule of thumb is that the higher interference (i.e., coherence) between sequences, the worse detection performance we may achieve. Following this principle, we partition the entire RM sequence space into multiple levels in Table~\ref{tab:RM_level} according to the coherence of the corresponding RM sequence space.
\begin{table}
  \centering
  \caption{Partition of RM sequence space by coherence}
    \begin{tabular}{|c|c|c|c|}
    \hline
    Space  & Space & \multirow{2}[2]{*}{Corresponding RM Sequences} & Max \bigstrut[t]\\
    Level & Size  &       & Coherence \bigstrut[b]\\
    \hline
    1     & $N = 2^m$ & Sequences generated by single $P$ & 0 \bigstrut\\
    \hline
    2     & $N^2$    & Seqs. generated by Kerdock set & $\frac 1 {\sqrt N}$ \bigstrut\\
    \hline
    3     & $N^3$    & Seqs. generated by $DG(m,1)$ set & $\frac{1}{N^{\frac{1}{2}-\frac{1}{m}}}$ \bigstrut\\
    \hline
    $\cdots$    & $\cdots$     & $\cdots$    & $\cdots$ \bigstrut\\
    \hline
    $l$     & $N^l$    & Seqs. gen. by $DG(m,l-2)$ set & $\frac{1}{N^{\frac{1}{2}-\frac{l-2}{m}}}$ \bigstrut\\
    \hline
    \end{tabular}%
  \label{tab:RM_level}%
\end{table}%

Our mapping from user ID to RM sequence in Section~\ref{section:RM} exactly follows Table~\ref{tab:RM_level}, where users with lower ID uses the sequences from lower-level space\footnote{Note that the higher-level space includes lower-level subspace.} with lower coherence. As we expand sequence space to achieve a lower collision rate, we should first use the RM sequences from lower-level subspace and then higher-level space. With this layered construction  of RM sequences, if we want to use $C$-sized sequence space, the optimal set of sequences are those with ID $\in \{0, \cdots C-1\}$ as described in Section~\ref{section:RM}. Through this layer-by-layer expansion, multi-user interference is mitigated to facilitate a better detection performance.

\section{Successive interference cancellation (SIC) based sequence detection}\label{section:SIC}
\subsection{Level-1 space}
The most elementary group of RM sequences are the RM sequences generated from a single $P$ matrix, i.e., level-1 space. The detection problem is the same as \eqref{y} except that $P_1 = \cdots = P_k = P$. To detect from \eqref{y} the active users, we simply perform a fast Walsh-Hardamard transform (FWHT) on the received signal $y$ to obtain the correlation values between $y$ and all $2^m$ possible transmitted signals $\phi_{P,b}$. The positions of the $k$ highest peaks correspond to the $b$ vectors of the $k$ active users. The above scenario is similar to the uplink grant-free multiple access discussed in \cite[Chap.~III-B]{SCMA:grant-free}, in which three algorithms with complexity $O(N^2)$, $O(N^2)$, and $O(N^3)$ are introduced, respectively. Here, leveraging the computationally efficient fast transform, the complexity is only $O(N \log N)$. Since all sequences in level-1 are orthogonal, they provide the best detection performance but have the minimal user space.

\subsection{Higher-level space}
When a larger user space is required, we can expand the user space by including the RM sequences generated from multiple $P$ matrices. In contrast to level-1 space, here the sequences of different users are no longer orthogonal. To suppress the inter-user interference as low as possible, successive interference cancellation (SIC) is adopted in \cite{RM:a-fast}. The SIC-based method is outlined in Algorithm~\ref{alg:SIC_outline}. The objective of the SIC-based algorithm is to recover the $P$ matrices of the $k$ users without traversing over all possible $P$ matrices. In each iteration, the algorithm extracts the $P$ matrix and $b$ vector associated with the highest-power user, and repeats this process until all users are detected.

\begin{algorithm}
\begin{algorithmic}
\STATE Input: $\{y\}$; Output: $\{h_l, P_l b_l, \forall l \in (1,\cdots,k)\}$
\STATE Initialization: $t=1$, $y_t = y$
\WHILE {$\{\parallel y_t\parallel > \epsilon \}$ or $t \leq t_{\max}$}
\STATE Extract from $y_t$ the $(\hat{P}_t,\hat{b}_t)$ of the largest-power user
\STATE Channel estimation for all the detected users:
\STATE $\qquad \arg \min_{\overrightarrow{c}}\parallel y-\sum_{l=1}^{t}\hat{h}_l \phi_{\hat{P}_l,\hat{b}_l}\parallel$
\STATE Cancel the detected sequence of the $t$ users:
\STATE $\qquad y_{t+1} = y-\sum_{l=1}^{t}\hat{h}_l \phi_{\hat{P}_l,\hat{b}_l}$
\STATE Increase $t$ by 1
\ENDWHILE
\end{algorithmic}
\caption{SIC-based sequence detection \cite{RM:a-fast}}
\label{alg:SIC_outline}
\end{algorithm}

The key step in Algorithm~\ref{alg:SIC_outline} is extracting the $P$ matrix of the largest-power user from the potentially huge user space. Once a $P$ is recovered, the corresponding $b$ vector can be recovered as in the level-1 space. According to Section~\ref{section:RM}, each $(P,b)$ pair uniquely determines a user ID.

\section{Improved RM detection for noise resilience}
The original recovery algorithm in \cite{RM:a-fast} works well in noiseless or high signal-to-noise ratio settings. However, its performance degrade quickly in wireless channels with both noise and fading. To combat the noisy channel in grant-free access, we modify the original algorithm and have harvested significant performance gain. The improvements are described as follows.

\subsection{Shuffling over multiple orders}
\begin{algorithm}
\begin{algorithmic}
\STATE Input: $\{y\}$; Output: $\{P,b\}$
\FOR {$p = 1 \to p_{\max}$}
\STATE $perm_p = {\tt randperm}(m)$.
\FOR {$j = perm_p(j)$}
\STATE Set unit weight $m$-length vector $e_j$, where the $j$-th element is 1, and all other elements are 0
\STATE Obtain $y(x+e_j)$ as follows: take the $y$ sequence, swap neighboring blocks of size $2^{j-1}$ (e.g., if $j=3$, then swap the values indexed by $1 \to 4$ with $5 \to 8$, and $9 \to 12$ with $13 \to 16$ and so on)
\STATE Pointwise multiply $y(x+e_j)$ with the conjugate of $y$
\STATE  $\qquad Y_{e_j} = {\tt fwht}(y(x+e_j).*{\tt conj}(y))$
\STATE The position of the highest peak of $Y_{e_j}$ corresponds to $\hat{P}_p$'s $j$-th column vector
\ENDFOR
\STATE Pointwise-multiply $y$ with the conjugate of $\phi_{P,b=0}$:
\STATE Take the fast Walsh-Hardamard transform:
\STATE $\qquad Y = {\tt fwht}(y.*{\tt conj}(\phi_{P,0}))$
\STATE Find the highest peaks of $Y$, take the binary form of the peak positions as $\hat{b}_p$
\ENDFOR
\STATE Compute the error metric for all candidates $(\hat{P}_p,\hat{b}_p)$, $p \in \{1 \cdots p_{\max}\}$.
\STATE Pick the $(P,b)$ pair with the lowest error metric as the next user
\end{algorithmic}
\caption{Shuffling based user detection}
\label{alg:SIC_shuffle}
\end{algorithm}
In Algorithm~\ref{alg:SIC_outline}, only one $\{P_t,b_t\}$ pair (one user) is extracted for interference cancellation in each iteration. If the $\{P_t,b_t\}$ pair is wrongly detected, canceling the signal associated with $\{P_t,b_t\}$ is equivalent to adding a fake user's sequence $c_t \phi_{P_t,b_t}$ to the received signal. This will not only cause false alarm but also incur more interference to the existing users. Therefore, we propose a two-fold strategy to avoid this situation. First, try to extract multiple $P$ candidates rather than only one $P$. Second, check the credibility of the extracted $\{P_t,b_t\}$ pairs before canceling its associated signal. The algorithm is described in Algorithm~\ref{alg:SIC_shuffle}.

According to \cite{RM:a-fast}, each $P$ matrix is detected based on the correlation between the received signal $y(x)$ and the shifted version of itself $y(x+e)$
\begin{equation}\label{chirpDecoding}
y(x+e)\overline{y(x)} = \frac {1} {2^m}\sum_{l=1}^k h_l^2 (-1)^{b_l^T e}(-1)^{e^T P_l x} + \texttt{chirps},
\end{equation}
where $e$ is a unit weight $m$-length binary vector where only the $j$-th bit is 1, and the chirps are multi-user interference and noise. Since the chirps are of lower power, applying FHWT to \eqref{chirpDecoding} results in peaks at position $P_l e$ which corresponds to the $j$-th column of the $P$ matrix. Repeating this $m$ times leads to the recovery of all $m$ columns of a $P$ matrix.

To extract the $P$ matrix in a more noise-resilient way, the first part of our strategy can be done by shuffling the orders of column recovery in Algorithm~\ref{alg:SIC_shuffle}. Instead of recovering the columns sequentially from the 1st to the $m$-th, we can try a different order each time. The orderings can be generated from random permutations of $[1,\cdots,m]$.
\begin{figure}
\centering
    \includegraphics[width= 0.47\textwidth]{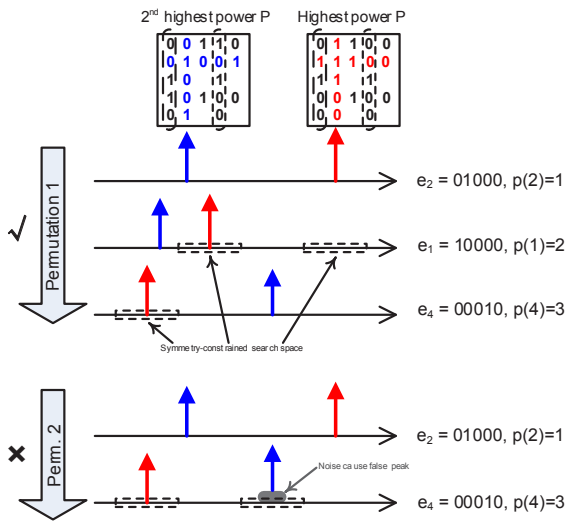}
    \caption{Shuffling intuition: an example that two different column-recovery orders may lead to extractions of different $P$ matrices. Permutation 1 is lucky to find the correct column vector, but Permutation 2 is deceived by noise. In practice, we may shuffle the order to obtain multiple $P$ candidates, and pick the most credible $P$ according to certain decision rule.}
    \label{fig:shuffle_illustration}
\end{figure}

We argue that different permutations may lead to extractions of different $P$. We assume that two users with similar received power have $P$ matrices as shown in Fig.~\ref{fig:shuffle_illustration}. Given that the 2nd column/row of the highest-power user has been correctly recovered, if we recover the 1st column in the next round (Permutation 1), we will not confuse with the 1st column of the 2nd highest-power user. This is because the search of the highest peak is conducted only in the symmetry-constrained space\footnote{All $P$ matrices are symmetric by definition.}, thus excluding the false peak. However, if we recover the 4-th column in the next round (Permutation 2), we may wrongly recover the 4-th column of the 2nd highest-power user. This time, the column vectors of both the highest-power user and the 2nd highest-power user are within the search space. The real peak may not be easily distinguished due to inter-user interference and noise.

The second part of our strategy help us to decide which of the extracted $\{P,b\}$ pairs is the most credible one. This can be done by exploiting some side information of the extracted $\{P,b\}$ pair. In the following, we propose two methods to compute the error score.
\subsubsection{Distance based decision}
Given two $m \times m$ matrices $P$ and $Q$, one way to tell which is more credible is the ranking of their column vectors in the Hadamard spectrum in terms of amplitude. If $P$ belongs to one of the accessing users, each column vector of $P$ should represents a peak equal or close to the highest peak. We sort the Hadamard spectrum of each column according to the amplitude in descending order. The distance of $P$'s $j$-th column vector $p_j$ is thus defined as the ranking of its peak in the Hadamard spectrum. Our error metric is the sum distance of all $m$ columns:
\begin{equation}
Dist(P) = \sum_{j=1}^m ranking(p_j).
\end{equation}
\begin{figure}
\centering
    \includegraphics[width= 0.47\textwidth]{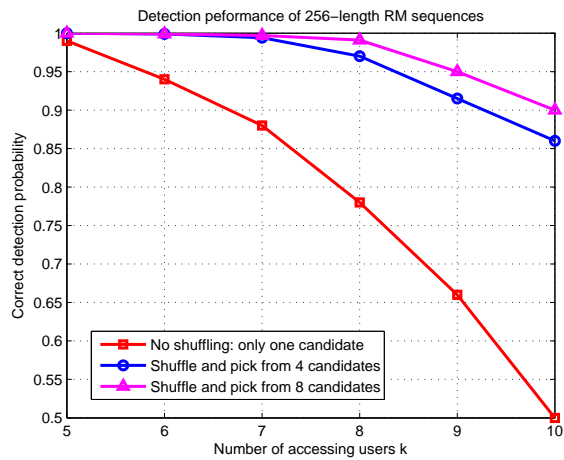}
    \caption{Shuffling gain: significant improvement is observed when we shuffle 4 times. As more shuffling is added, the improvement diminishes.}
    \label{fig:shuffle_gain}
\end{figure}

We pick the lowest-sum-distance $\hat{P}_p$ and its associated $\hat{b}_p$ from all candidates as the extracted $(P,b)$ pair.

\subsubsection{Residual energy based decision}
A more powerful but more computationally complex method is selecting $\{P,b\}$ pair according to the residual energy after signal cancellation. We perform channel estimation for all sequences generated by the $\{P,b\}$ candidates, and calculate their residual energy by canceling  $c_t \phi_{P_t,b_t}$. Intuitively, if the $\{P,b\}$ being tested is from a real user, then residual energy should decrease after its signal is canceled. Therefore, we may use the residual energy as the error metric, and pick the $\{P,b\}$ pair with the lowest residual energy as the extracted $P$.

Shuffling can also dramatically improve the detection performance in certain scenarios. For $m=8$, the detection results for ``no shuffling, shuffle and pick from 4 $\{P,b\}$ candidates, shuffle and pick from 8 $\{P,b\}$ candidates'' are shown in Fig.~\ref{fig:shuffle_gain}. We observe significant improvement in the detection performance even if a few candidates are generated for decision. Note that the fast Hadamard transform of all $m$ columns is computed only once at the beginning, therefore shuffling itself does not incur additional complexity. However, if we use the residual energy based decision rule, additional computation is incurred during additional channel estimation.

The complexity of Algorithm~\ref{alg:SIC_shuffle} is $O(m(m+1)2^m)$, where the main computation tasks are the H-transforms. Algorithm~\ref{alg:SIC_outline} iterate $k$ times to detect all $k$ users, the overall complexity is thus $O(k \times m(m+1)2^m)$. Note that this complexity is not dependent on the value of $r$, therefore also independent of the user space size. In contrast, traversing all sequences yields complexity $O(2^{m(r+2)}\times 2^m)$.

\subsection{Double checking the detected $P$}
Sometimes the recovered $P$ candidates may violate the rank property, i.e., $rank(P) \geq m-2r$ \cite{RM:construction}. These candidates are absolutely wrong and should be discarded. Thanks to the shuffling technique, we may re-permute the column recovery order until we obtain a valid $P$.

Occasionally an already detected user can be re-detected in subsequent iterations. This will induce an endless loop if not properly intervened. In this case, we may shuffle the column recovery order to extract a distinct $\{P,b\}$ pair. In the worst case, we keep shuffling until we find something new.

\section{Link-level simulation results}
To validate the proposed sequence design and the improved detection algorithm, we conducted extensive experiments using our SCMA-based grant-free massive access simulator. Two potential configurations for 5G are examined, i.e., narrow-band IoT for massive connectivity and wideband OFDM for ultra-reliable and low-latency communications.
\subsection{Narrowband IoT}
First, we examine the performance of RM sequences as preamble in the narrow-band massive connectivity setting. The simulation parameters (e.g., bandwidth) follow the most recent 3GPP draft report \cite{3GPP:RAN1} which describes the possible future IoT configurations, and are listed in Table~\ref{tab:NB}.
\begin{table}[htbp]
  \centering
  \caption{Narrow Band IoT Configuration}
    \begin{tabular}{|c|c|}
    \hline
    Channel & EPA Urban Micro (3Km/h) \bigstrut\\
    \hline
    \# Subcarrirers & 4 \bigstrut\\
    \hline
    Subcarrier Spacing & 3.75 KHz \bigstrut\\
    \hline
    Antenna & 1 Tx, 2 Rx \bigstrut\\
    \hline
    Preamble Power & Same as active data tone \bigstrut\\
    \hline
    Subcarrier per Preamble & 1 \bigstrut\\
    \hline
    Preamble Length & 64 symbols \bigstrut\\
    \hline
    Uplink Frame Length & 80 ms \bigstrut\\
    \hline
    Contention-based Access & Yes \bigstrut\\
    \hline
    \end{tabular}%
  \label{tab:NB}%
\end{table}%

In the narrow-band setting, we restrict our attention on the detection \& collision performance under massive connectivity. We assume that each IoT device transmits once per hour, which corresponds to an active probability of $9 \times10^{-7}$. To support massive connectivity in 5G IoT, a base station that covers 1 km$^2$ should serve about $10^7$ links \cite{5G:features}. That means about 5 to 10 users will be active at the same time. Therefore, we simulated two typical cases, 10 users simultaneously access under medium SNR (7dB), and five users simultaneously access under relatively low SNR (1dB). In our detection algorithm, we pick the $k$ users with highest energy using the proposed algorithm, where $k$ equals the number of active users. In each grant-free access, each user randomly chooses an ID from $[0,\cdots,C-1]$, where $C$ is the sequence space size, and uses the corresponding RM sequence as its preamble. When more than one user chooses the same sequence, we count them as collisions. If our algorithm fails to detect an active user, we count it as a miss detection. Both collisions and miss detection lead to packet losses. Therefore we also count the sum of collisions and miss detection as total failures.

\begin{figure}
\centering
    \includegraphics[width= 0.47\textwidth]{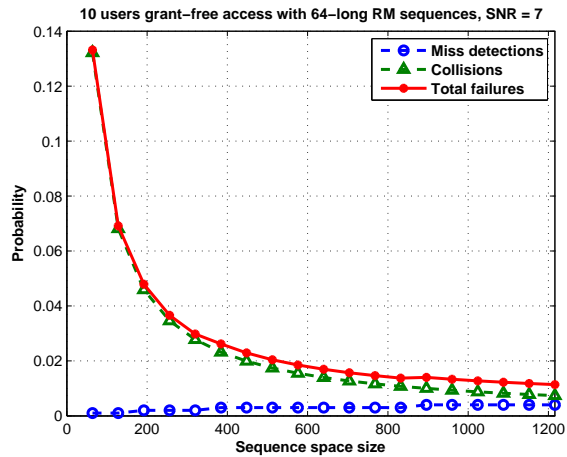}
    \caption{The total access failures drops quickly thanks to the huge sequence space provided by RM.}
    \label{fig:nb_10_ue_snr_7}
\end{figure}

\begin{figure}
\centering
    \includegraphics[width= 0.47\textwidth]{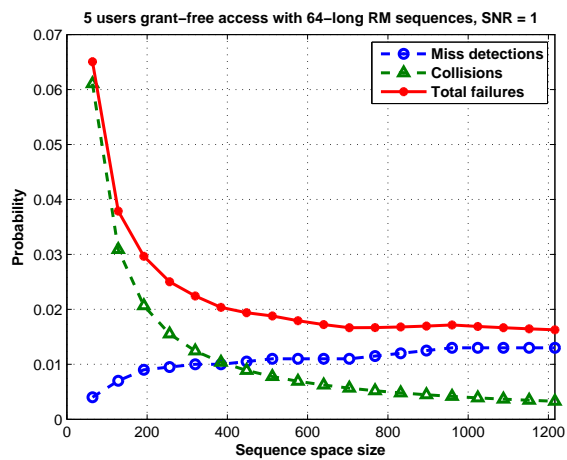}
    \caption{Even in very adverse channel condition, the miss detection rate only slightly grows thanks to our robust algorithm.}
    \label{fig:nb_5_ue_snr_1}
\end{figure}

The results of these three metrics are given in Fig.~\ref{fig:nb_10_ue_snr_7} and Fig.~\ref{fig:nb_5_ue_snr_1}, respectively. The observation is as follows. On the one hand, thanks to the huge sequence size provided by RM, the number of collisions between users drops quickly as the sequence space expands. On the other hand, facilitated by our detection algorithm, the miss detection rate only grows slightly with the space size even under low SNR, which exhibits robustness under multi-user interference and noise. In summary, we conclude that it is worthwhile to dramatically expand the user sequence space in 5G IoT scenarios with massive connections.


\subsection{Wideband OFDM-SCMA}
We use RM sequences as pilots for a OFDMA-based grant-free access system to realize ultra-reliable and low-latency communications (e.g., one-shot ($\leq 1$ms) transmission with $\leq 10^{-4}$ packet error rate). The simulation configuration is compatible with the incumbent LTE system and is listed in Table~\ref{tab:WB}. This time we not only examine the collision \& detection performance, but also include data transmission performance. Based on OFDM, we also implement sparse code multiple access (SCMA) to obtain a 3x spectrum efficiency (interested readers are referred to \cite{SCMA:seminal} for details.). Note that our pilot design is not restricted to SCMA, but also applies to other modulation techniques. Apart from active user detection, the pilot sequences are also used for multi-path channel estimation. Here we use the SCME Urban Macro channel model, which has relatively long delay spread than the EPA channel. In such a case, MMSE multi-user channel estimation \cite{CE:MMSE} is adopted to address the severe multi-path fading.
\begin{table}
  \centering
  \caption{Wideband OFDM-SCMA Configuration}
    \begin{tabular}{|c|c|}
    \hline
    Channel & SCME Urban Macro (3Km/h) \bigstrut\\
    \hline
    \# Total Subcarriers & 1024 \bigstrut\\
    \hline
    \# Subcarriers Used & 72 \bigstrut\\
    \hline
    Subcarrier Spacing & 15 KHz \bigstrut\\
    \hline
    Antenna & 1 Tx, 2 Rx \bigstrut\\
    \hline
    Pilot Power & Same as active data tone \bigstrut\\
    \hline
    Subcarriers per Pilot & 72 \bigstrut\\
    \hline
    \# Pilot Length & 2 OFDM symbols \bigstrut\\
    \hline
    \# Uplink Frame Length & $0.8$ ms \bigstrut\\
    \hline
    Contention-based access & Yes \bigstrut\\
    \hline
    SCMA spreading factor & 4 \bigstrut\\
    \hline
    Channel Code & Turbo 1/3 (including CRC) \bigstrut\\
    \hline
    \end{tabular}%
  \label{tab:WB}%
\end{table}%

We compare the performances between the RM-based scheme and the ZC-based scheme used in the incumbent LTE system. The sequence design of the two schemes is as follows. The RM sequences have length 127 and the ZC sequences have length 139. An uplink frame has 12 OFDM symbols. For both sequences, we perform a 144-point discrete Fourier transform (DFT) and map them onto the 144 subcarriers on the 4-th and the 11-th symbols. For ZC sequences, we use 12 roots and 12 cyclic shifts for each root; therefore we have 144 ZC pilots in total. For RM sequences, we set the sequence space $C=16000$ and use the sequences from ID 0 to 15999 according to Section~\ref{section:RM}; therefore we have 16000 RM pilots in total.

\begin{figure}
\centering
    \includegraphics[width= 0.47\textwidth]{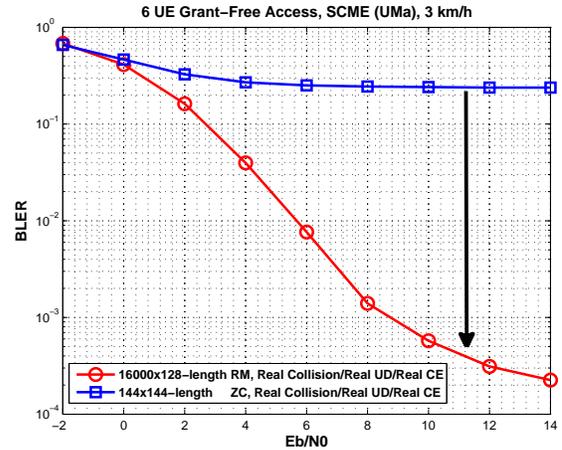}
    \caption{The proposed RM-based grant-free access scheme has significant gain over the traditional ZC-based scheme.}
    \label{fig:WB6UE}
\end{figure}

As shown in Fig.\ref{fig:WB6UE}, we let six active users simultaneously access by randomly choosing a sequence, and examine (i) collision, (ii) detection and (iii) block error rate (BLER). Due to the dramatically increased pilot space (from 144 for ZC to 16000 for RM), the collision rate reduces from 0.0342 to $10^{-4}$. Since the collision rate is independent from SNR, it will impose an error floor to the grant-free system. The collision-incurred performance bottleneck may be greatly loosened by the huge space of RM sequences, while the error floor is high for the ZC-based scheme. With our detection algorithm, our miss detection rate can be as low as $10^{-5}$ in the high SNR regime. Finally, combining all aspects, the block error rate (BLER) performance shows that RM-based pilot design achieves a packet loss rate close to $10^{-4}$  \textit{without} any scheduling and re-transmission. Since the packet length is 12 OFDM symbols which last about $0.8$ms, the ultra-reliable and low-latency requirements are satisfied.

\section{Conclusion}
In this work, we revealed the potential of RM sequences for user identification in grant-free massive access. While the collision rate can be reduced by sequence space expansion, the sequence construction and detection algorithm need to be re-designed to address the noise and multi-user interference in wireless channel. As an initial study, we implemented our scheme in narrow-band and wideband grant-free access systems. The former validates the massive connectivity scenario; the latter validates the ultra-reliable and low-latency communications scenario. In both cases, RM sequences helped to fulfill the harsh requirements of 5G.

\ifCLASSOPTIONcaptionsoff
  \newpage
\fi

\end{document}